\documentclass[a4paper,11pt]{article}
\pdfoutput=1 

\usepackage{jcappub} 

\usepackage[T1]{fontenc} 

\usepackage{color}
\usepackage[usenames, dvipsnames]{xcolor}
\usepackage{graphicx}	
\usepackage{amssymb}	

\makeindex

\title{An Optimally Weighted Estimator of the Linear Power Spectrum Disentangling the Growth of Density Perturbations Across Galaxy Surveys}

\author[a,b]{D. Sorini}

\affiliation[a]{Max-Planck-Institut f\"ur Astronomie, K\"onigstuhl 17, D-69117 Heidelberg, Germany}
\affiliation[b]{Imperial Centre for Inference and Cosmology, Imperial College, Blackett Laboratory, Prince Consort Road, London SW7 2AZ, United Kingdom}

\emailAdd{sorini@mpia-hd.mpg.de}

\abstract{Measuring the clustering of galaxies from surveys allows us to estimate the power spectrum of matter density fluctuations, thus constraining cosmological models. This requires careful modelling of observational effects to avoid misinterpretation of data. In particular, signals coming from different distances encode information from different epochs. This is known as ``light-cone effect'' and is going to have a higher impact as upcoming galaxy surveys probe larger redshift ranges. Generalising the method by Feldman et al. (1994) \cite{FKP}, I define a minimum-variance estimator of the linear power spectrum at a fixed time, properly taking into account the light-cone effect. An analytic expression for the estimator is provided, and that is consistent with the findings of previous works in the literature. I test the method within the context of the halo model, assuming the cosmological parameters given by Planck Collaboration et al. (2014) \cite{Planck_2014}. I show that the estimator presented recovers the fiducial linear power spectrum at present time within 5\% accuracy up to $k \sim 0.80\;h\,\mathrm{Mpc}^{-1}$ and within 10\% up to $k \sim 0.94\;h\,\mathrm{Mpc}^{-1}$, well into the non-linear regime of the growth of density perturbations. As such, the method could be useful in the analysis of the data from future large-scale surveys, like Euclid.}

\begin{document}
\maketitle
\flushbottom


\section{Introduction} 
\label{Introduction}
Over the last few decades, more and more precise measurements of the Cosmic Microwave Background (CMB) have shown that the Universe was almost isotropic at the epoch of last scattering, presenting fluctuations in the temperature of the CMB of order $10^{-4}\; \mathrm{K}$ ~\cite{Bennett_2013, Planck_2014, Planck_2015}.  This remarkable discovery has led to the description of the Universe as initially homogeneous, except for tiny perturbations in the density of matter, which were set by inflation. Such seed fluctuations have then grown due to gravitational instability, eventually giving rise to the large-scale structure observable today. 

The evolution of photons, dark matter and baryons is governed by the Einstein-Boltzmann equations. As long as matter density perturbations are small, they can be linearised around the solution relative to a perfectly homogeneous and isotropic Universe. In this regime the growth of fluctuations is relatively straightforward, since their amplitude is simply enhanced by a time-dependent growth function, set by the cosmological parameters. If the fluctuations are not much smaller than the mean density, one needs to perturb Einstein-Boltzmann equations to higher orders, or to abandon perturbation theory in favour of semianalytic or numerical methods.

Within or outside the regime of applicability of linear theory, the choice of the cosmological model impacts the observed large-scale structure. So, by observing structures today, one can infer how likely a certain model is, and constrain its parameters. Galaxy surveys play a major role in this respect. The distribution of galaxies is thought to be a biased tracer of matter and its number density is described as a Poisson sampling of the underlying matter distribution. Consequently, mapping the galaxy distribution tells us also how matter is distributed. More importantly, the statistical properties of galaxy clustering represent a formidable source of information. 

Two widely used statistics are the two-point correlation function and the power spectrum. They are essentially the same statistics, expressed in real and Fourier space, respectively. In fact, the correlation function is defined as
\begin{equation}
\label{eq:correlation_function}
	\xi (\textbf{r}-\textbf{r}') = \langle \delta _m (\textbf{r}) \delta _m(\textbf{r}') \rangle ,
\end{equation}
where $\delta_m(\textbf{x})$ is the density fluctuations field. The power spectrum is nothing but its Fourier transform\footnote{In equations \eqref{eq:correlation_function} and \eqref{eq:power_spectrum} the large-scale homogeneity of the matter density field has been implicitly assumed.}
\begin{equation}
\label{eq:power_spectrum}
	P(\textbf{k}) = \int d^3r\,e^{i \textbf{k} \cdot (\textbf{r}-\textbf{r}')} \xi(\textbf{r}-\textbf{r}') \, .
\end{equation}
Both  correlation function and  power spectrum are affected by cosmological parameters. This is the reason why these functions are so important in cosmological studies. They can be predicted from observations and then used to constrain models.

A systematic study of the statistical properties of clustering was initiated by \cite{Yu_1969, Peebles_1973} and \cite{Peebles_1974}. These pioneering works informed an entire research field, focusing on the understanding of the two-point correlation function (e.g. \cite{Efstathiou_1991, Neuschaefer_1991, Couch_1993, Cole_1994, Matarrese_1997, Bagla_1998, Nishioka_1999, Hamana_2001, Moscardini_2002, Fedeli_2009, Raccanelli_2014, Clerkin_2015}, and references therein). The two-point correlation function contains full information if the distribution of the amplitudes of the primordial perturbations is Gaussian, otherwise it contains only partial information, but still represents a good starting point. Formally, studying the power spectrum instead of the two-point correlation function is absolutely equivalent, the former being the Fourier transform of the latter. However, this is not necessarily true when considering the estimates of these two quantities, which are obtained from finite and noisy observational data. The statistical analysis of galaxy surveys through the power spectrum represented indeed a flourishing area of research (e.g. ~\cite{Baumgart_1991, Vogeley_1992, Kaiser_1991, Strauss_1990, Fisher_1993, Webster_1977, Peacock_1991, Peacock_1992, Einasto, Park_1992, da_Costa_1994, Meiksin_1999, Hamilton_2000, Nishioka_2000, Munoz_2008}, and references therein). An advantage of studying the power spectrum with respect to the two-point correlation function is that it is always positive, so it is easier to detect any non-physical result. Besides, its shape does not depend on the knowledge of the mean number density of galaxies, which affects the $k=0$ mode only. Also, the power spectrum is the quantity which is generally provided by theories studying the early Universe (e.g. inflation). ~\cite{FKP}

It is of utmost importance to develop solid methods to infer the power spectrum from galaxy surveys. A fundamental work in this respect was published by \cite{FKP}, who defined an estimator of the power spectrum of galaxy clustering, given a certain galaxy distribution. The estimator is defined such that its variance divided by the power spectrum is minimised. I shall refer to ~\cite{FKP} as ``FKP'' throughout this work.

The FKP method is subject to some approximations and limitations. In particular, with galaxy surveys becoming wider and deeper, providing the correct estimate of such quantity is not trivial at all. One must indeed take into account various observational effects, like the redshift evolution of the luminosity function clustering amplitude and bias
(e.g. ~\cite{Matarrese_1997, deLaix_1998, Yamamoto_1999, Nishioka_2000, PVP}), the contribution of modes larger than the volume of the survey (e.g. ~\cite{de_Putter_2012}), and general-relativistic effects (e.g. ~\cite{Yoo_2009, Yoo_2010}). Various extensions of FKP have been realised, incorporating for example the light-cone effect ~\cite{Yamamoto_2003} and redshift space distortions ~\cite{Yamamoto_2005}. More sophisticated analyses followed, based for example on non-Gaussian contributions to likelihood analysis (e.g. ~\cite{Takahashi_2001}), Bayesian or hybrid approaches (e.g. ~\cite{Efstathiou_2004, Jasche_2010}), N-body simulations (e.g. ~\cite{Colombi_2009}), and multipole-analysis of the power spectrum ~\cite{Sato_2013, Kanemaru_2015}. 

In this paper, I shall focus on the light-cone effect. With this term I mean the fact that one is able to observe only one's own past light-cone. So, if the survey is particularly deep, one may receive signals coming from epochs spanning a significant lapse of time. Consequently, one is not really able to map the galaxy distribution, and thus compute its power spectrum, at a certain fixed time. A possible way to deal with this issue is looking for an estimator of the power spectrum, somehow averaged over a certain time lapse. This approach was adopted by \cite{Yamamoto_2003}, who was able to generalise the FKP method including, among others, the light-cone effect.

In the current paper, a different approach is attempted. I define an estimator of the power spectrum at a certain fixed time, including the light-cone effect in the FKP method, under certain assumptions on the behaviour of the correlation function. Section  \ref{FKP_light_cone} contains the main result of this work. In that section, I present my generalisation of FKP, and in section \ref{sec:implementation} how I implemented it in a test-case to determine its accuracy. In sections \ref{Results} and \ref{Discussion} I present and discuss the results, respectively. In section \ref{Conclusions} I state my conclusions and give an overview on the possible perspectives of this work. Unless otherwise indicated, all units mentioned in the text are comoving.

\section{Description of the Method}
\label{FKP_light_cone}

In this section I generalise the FKP method, considering the growth of density perturbations across a galaxy survey. To do that, I will follow the logic and notation of the original FKP paper. In section \ref{sec:intro_growth} I introduce the formalism necessary to address the problem, and show that, within linear perturbation theory, generalsing FKP is relatively straightforward. There is no original result in  subsection \ref{sec:intro_growth}. It serves only as an introduction to the subsequent  subsections, in which I consider non-linear growth of pertrubations. That is the most interesting case, since the applicability of linear theory is limited to large scales only. The method described in subsections \ref{sec:sub_2}-\ref{sec:final_weighting} represents original work.

\subsection{Growth of Perturbations}
\label{sec:intro_growth}

When performing observations, one actually receives light signals coming from a range of distances and thus from different epochs. The measured overdensities in two points $\textbf{r}_1$ and $\textbf{r}_2$ are then $\delta (\textbf{r}_1, \, t_1)$ and $\delta (\textbf{r}_2, \, t_2)$. In contrast, one is interested in determining $\delta (\textbf{r}, \, t_0)$, i.e. the overdensity field at different positions, at the same reference time $t_0$ (for example, today). It is then necessary to develop a formalism to connect the overdensity at the generic time $t$ to the overdensity at $t_0$. 

Making the time-dependence of the overdensity field explicit, the number density of galaxies in a survey is given by
\begin{equation}
\label{eq:nr_generic}
	n_g(\textbf{r}, \, t) = \bar{n} (\textbf{r}, \, t) [1+ \delta (\textbf{r}, \, t)] \, ,
\end{equation}
where $\bar{n}(\textbf{r}, \, t)$ is the expected number density of galaxies, given the luminosity and angular selection criteria of the survey. Since time is a function of the distance $r$, the explicit time dependence in equation \eqref{eq:nr_generic} may be suppressed. Within linear theory, equation \eqref{eq:nr_generic} can be written as
\begin{equation}
\label{eq:nr_linear}
n _{g} (\textbf{r}) = \bar{n} (\textbf{r}) \left[1+ \frac{D(r)}{D(r_0)} \delta _0 (\textbf{r}) \right] \, ,
\end{equation}
where $D(r)$ is the growth function \cite{Smith_2003}, and $\delta _0 (\textbf{r})$ is the overdensity at the reference time $t_0=t(r_0)$. I define the corresponding weighted galaxy fluctuation field as
\begin{equation}
\label{eq:Fk_def}
	F( \textbf{k} ) = \frac{\int d^3r \, e^{i \textbf{k} \cdot \textbf{r}} w(\textbf{r}) [n_{g}(\textbf{r})- \alpha n_s(\textbf{r})]}{\int d^3r \, \left(\frac{D(r)}{D(r_0)}\right)^2 w(\textbf{r})^2 \bar{n}(\textbf{r})^2} \, ,
\end{equation}
where $w(\textbf{r})$ the weight function and $n_s (\textbf{r})$ is the synthetic catalogue. The constant $\alpha$ is a rescaling factor to normalise the total number of galaxies in the synthetic catalogue to the real catalogue. Since the synthetic catalogue has in general many more galaxies than the real one, $\alpha$ is usually small. In the limit of infinitely many synthetic galaxies, $\alpha$ tends to zero.

The optimal estimator of the linear power spectrum at fixed time can be immediately obtained noting the analogy with the work published by Percival, Verde and Peacock in 2004 (hereafter, PVP) \cite{PVP}. Their work focused on an extension of the FKP method considering a set of galaxies of a certain luminosity, from a Poisson sampling of a linearly biased density field. In PVP, the density of galaxies with luminosity $L$ is given by 
\begin{equation} \label{eq:density_PVP}
	n_g(\textbf{r},\,L) = \bar{n}(\textbf{r},\,L) [1+b(\textbf{r},\,L)\delta(\textbf{r})]\, ,
\end{equation}
where $b(\textbf{r},L)$ is the bias. Comparing the definition of the galaxy density in equations \eqref{eq:density_PVP} and \eqref{eq:nr_linear}, it is clear that the linear growth of density perturbations can be formally treated as a linear bias. Therefore, it is sufficient to follow PVP replacing $b(\textbf{r},\,L)$ with $D(r)/D(r_0)$. Following FKP, the estimator $\widehat{P}_0(\textbf{k})$ of the linear power spectrum at fixed time $t_0$ is defined as
\begin{equation}
	\widehat{P}_0(\textbf{k}) =\langle  \vert F(\textbf{k}) \vert ^2 \rangle -  P_{\textrm{shot}} \, ,
\end{equation}
where $P_{\textrm{shot}}$ is the shot noise. The resulting optimal weight function is
\begin{equation}
\label{eq:weight_func_linear}
	w_0(\textbf{r}, k) = \frac{P_0(k)}{1 + \left(\frac{D(r)}{D(r_0)}\right)^2 \bar{n}(\textbf{r}) P_0(k)} \, .
\end{equation}
Equation \eqref{eq:weight_func_linear} is consistent with the expression found in PVP. They are not exactly the same because I adopted a different definition of the galaxy density fluctuations in \eqref{eq:Fk_def}. In the context of this manuscript, that definition is preferable, since it allows for a more straightforward comparison with the main results of this work, which are presented in subsection \ref{sec:final_weighting}. Nevertheless, it can be verified that, adopting the same definition of $F(\textbf{k})$ as in PVP, the expressions for the optimal weight function are fully consistent.

Assuming linear growth of perturbations makes the estimate of the fixed-time linear power spectrum from galaxy surveys particularly straightforward. However, this assumption restricts the validity of the method at large scales only. It is then interesting to develop a formalism to estimate the power spectrum at fixed time, considering non-linear growth of perturbations across the galaxy survey. In the following subsections, I shall generalise the FKP method to the mildly non-linear regime. I shall call $P(k)$ the fully, non-linear power spectrum, and $\Delta^2(k)=k^3P(k)/2\pi^2$ the dimensionless non-linear power spectrum. In the remainder of the paper, I shall denote with $\Delta ^2 _{\textrm{L}} (k)$ the dimensionless power spectrum predicted by linear theory. When referring to $\Delta^2(k)$ and  $\Delta ^2 _{\textrm{L}} (k)$, I shall often omit ``dimensionless''. That should not be confusing, as it should be clear from the context whether I am referring to $P(k)$ or $\Delta ^2 (k)$.

There have been various attempts to get an approximate analytic expression for non-linear power spectrum. The most straightforward way to accomplish that is perhaps perturbation theory, where the equations for the evolution of the matter density field are perturbed beyond first order (see \cite{Bernardeau_2002} for a review). A different idea to address the issue is expressing the non-linear power spectrum (or correlation function) as a function of its linear counterpart, evaluated at a different scale \cite{Hamilton_1991, Peacock_1994, Peacock_1996, Padmanabhan_1996, Jain_1997}. Another approach is assuming a power law for the shape of the  primordial power spectrum and the time evolution of the scale factor, and then determine the time evolution of the scale at which perturbations become non linear. It can be shown that, under certain conditions, the power spectrum evolves in time according to a self-similar solution \cite{Davis_1977, Efstathiou_1988, Colombi_1996, Jain_1998}. A different widely used analytic method to describe non-linear clustering is the ``halo model'' (\cite{Neyman_1952, Scherrer_1991, Seljak_2000, Ma_2000, Peacock_2000, Scoccimarro_2011}; see \cite{Cooray_2002} for a review). The basic idea consists in splitting the power spectrum into two terms. One of them, the ``two-halo'' term, models large-scale correlations in the density of matter residing in different halos. The other one, the ``one-halo'' term, represents the contribution to the power spectrum due to correlations of the density field within the same halo. The parameters of the analytic functional form of the power spectrum provided by the halo model can be determined fitting results of N-body cosmological simulations \cite{Smith_2003}. This semi-analytic method is called ``halofit'' model \cite{Takahashi_2012}. In this work, I will consider the revised version of the halo model provided by \cite{Takahashi_2012}.

\subsection{Revised Halofit Model}
\label{sec:sub_2}
I shall now briefly review the revised halofit model by \cite{Takahashi_2012}. Let us define
\begin{equation}
\label{eq:two-halo-P}
	\Delta^2(k)=\Delta^2_{\textrm{Q}}(k)+\Delta^2_{\textrm{H}}(k) \, ,
\end{equation}
where $\Delta^2_{\textrm{Q}}(k)$ and $\Delta^2_{\textrm{H}}(k)$ are the two-halo and one-halo terms, respectively. Unlike in linear theory, they cannot be expressed as a product of a function of $k$ only and a function of time (or, equivalently, distance or redshift) only. Specifically,
\begin{align}
\label{eq:Takahashi_DQ}
	\Delta^2_{\textrm{Q}}(k) &=\Delta^2_{\textrm{L}}(k) \left\lbrace \frac{[1+\Delta^2_{\textrm{L}}(k)]^{\beta_n}}{1+\alpha_n \Delta^2_{\textrm{L}}(k)}\right\rbrace e^{-B(y)}\\
	\Delta^2_{\textrm{H}}(k) &=\frac{1}{1+\mu_n y^{-1}+\nu_n y^{-2}} \cdot \frac{a_n y^{3 f_1(\Omega_m)}}{1+b_n y^{f_2(\Omega_m)}+[c_n f_3(\Omega_m)y]^{3-\gamma_n}} \, ,
\end{align}
where
\begin{equation}
	B(y) =\frac{y}{4}+\frac{y^2}{8}\, .
\end{equation}
In the equation above, $y=k/k_{\sigma}$, with $k_{\sigma}$ defined as $\sigma^2(k^{-1}_{\sigma}) =1$, where 
\begin{equation}
	\sigma^2(R) =\int d\ln k \; \Delta^2_{\textrm{L}}(k) e^{-k^2 R^2} \, .
\end{equation}
The effective spectral index $n_{eff}$ and the curvature $C$ are defined as
\begin{align*}
	n_{eff}+3 &=-\left. \frac{d \ln \sigma^2(R)}{d \ln R} \right\vert_{\sigma=1}\\
	C &=-\left. \frac{d^2 \ln \sigma^2(R)}{d \ln R^2} \right\vert_{\sigma=1} .
\end{align*}
All other parameters are determined as polynomial functions of $n_{eff}$ and $C$. The coefficients of the polynomials are obtained by fitting results of N-body simulations. The details can be found in \cite{Takahashi_2012}.

\begin{figure}
	\includegraphics[width=\columnwidth]{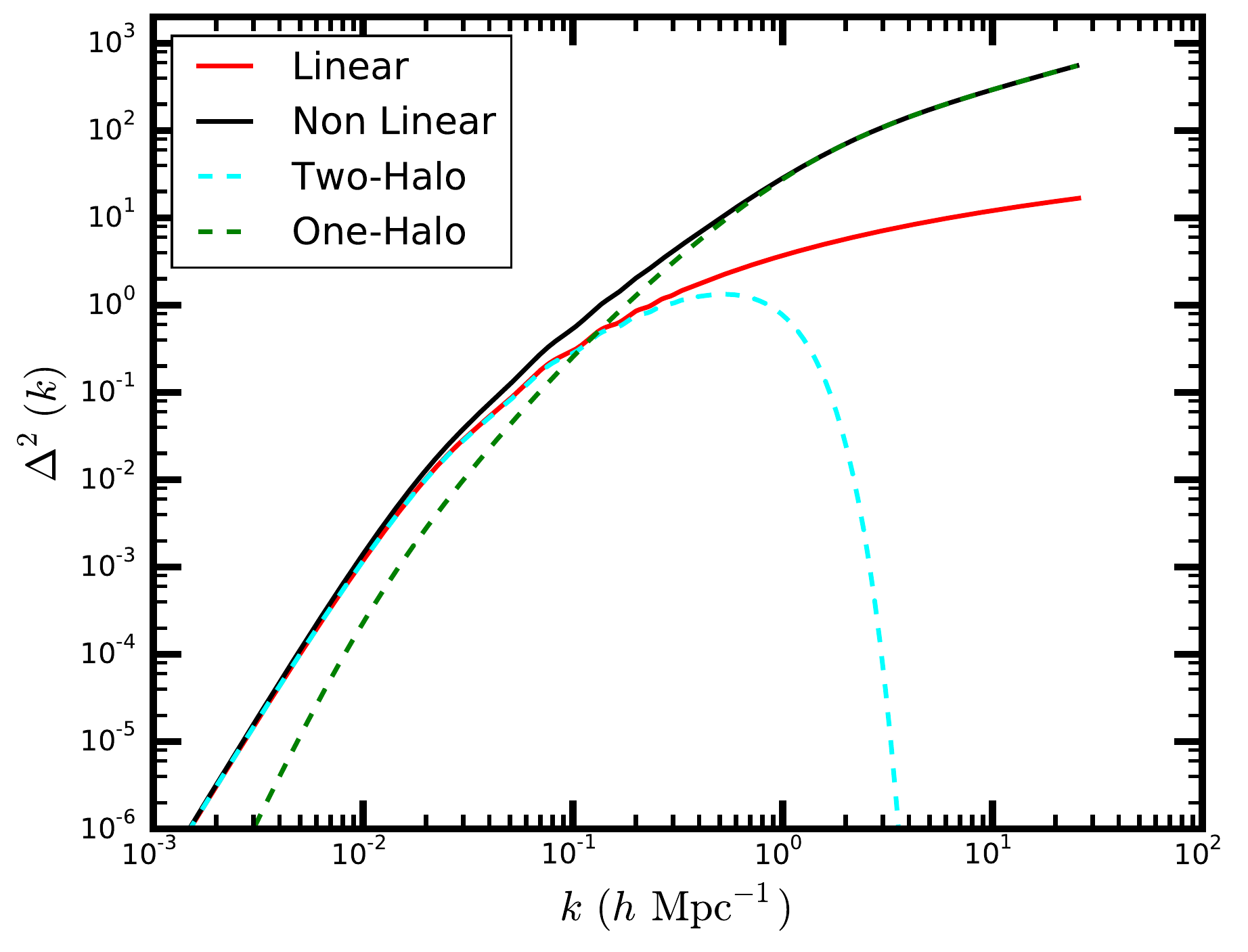}
	\caption{Two-halo model adopted in this work, relative to Planck 2014 cosmology. The solid red line is the linear power spectrum at redshift $z=0$. The dashed cyan and green lines are the corresponding two-halo and one-halo terms, respectively. The non linear power spectrum at $z=0$ is represented by the solid black curve. Within the context of the two-halo model, it is given by the sum of the two-halo and one-halo terms.}\label{delta L and delta Q}
\end{figure}

The halofit model can be visualised in figure \ref{delta L and delta Q}. The solid red line is the linear power spectrum generated through CAMB (~\cite{CAMB_1, CAMB_2}) with the cosmological parameters given by ~\cite{Planck_2014} ($\Omega_{\mathrm{m}}=0.315$, $\Omega_{\Lambda}=1-\Omega_{\mathrm{m}}=0.683$, $\Omega_{\mathrm{b}}=0.049$, $h=0.673$, $A_s=2.215 \times 10^{-9}$, $n_s=0.96$). The dashed cyan and green lines are the corresponding two-halo and one-halo terms, respectively, with the parametrisation given by \cite{Takahashi_2012}. The solid black line is the sum of the two terms, i.e. the non-linear power spectrum. One can clearly see that the two-halo term dominates at scales $k \lesssim 0.02\, h \,\rm Mpc^{-1}$, where the power spectrum is very well described by linear theory. Moreover, the two-halo term approximates the linear power spectrum up to $k \sim 0.2 \, h \, \rm Mpc^{-1}$. On the contrary, for $k \gtrsim 0.6 \, h \,\rm Mpc^{-1}$, the power spectrum strongly deviates from linearity and is very well approximated by the one-halo term. In the range of modes between these two limits it is necessary to consider both terms for an accurate description of the non-linear power spectrum. 

\begin{figure*}
	\includegraphics[width=0.5\columnwidth]{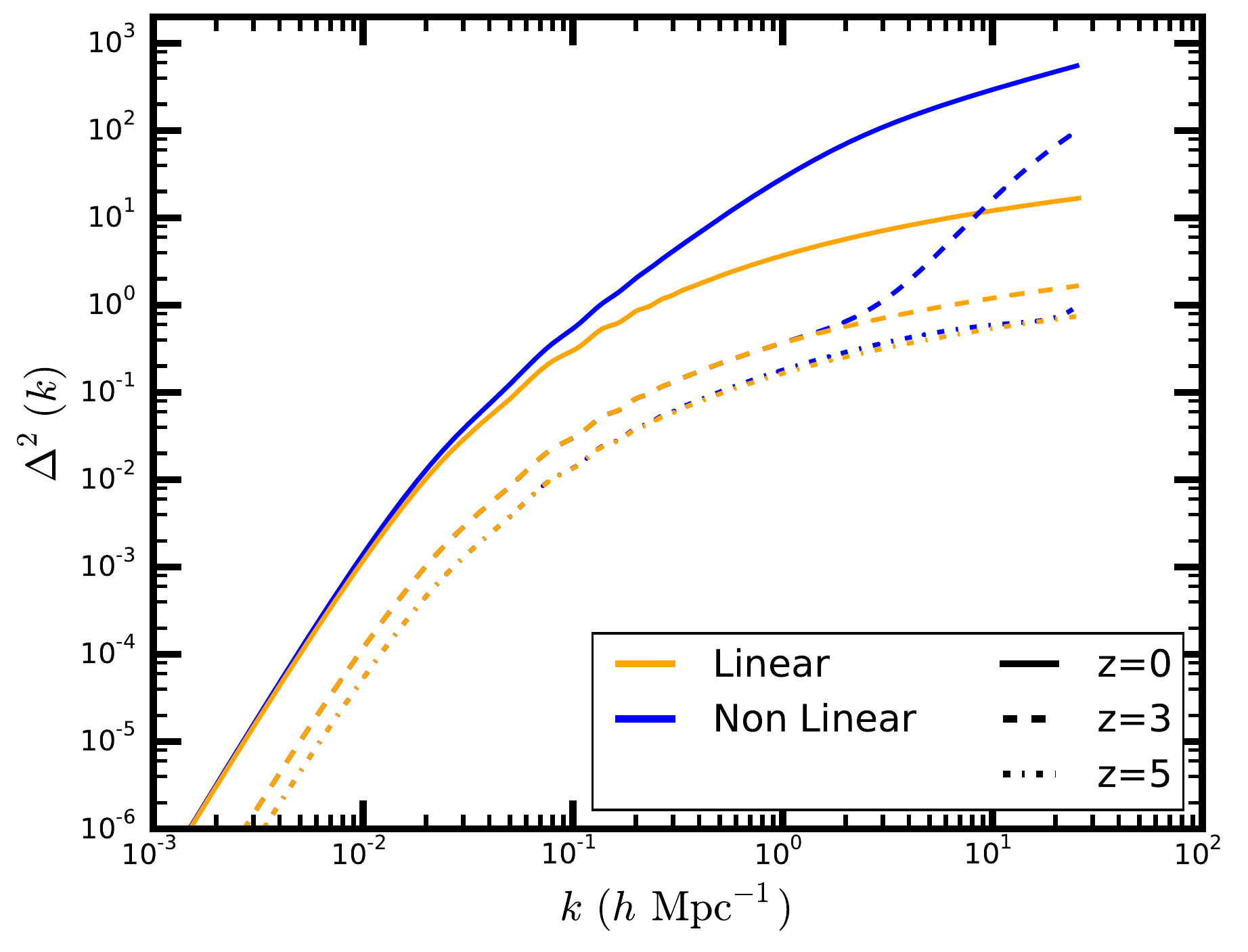}
	\includegraphics[width=0.5\columnwidth]{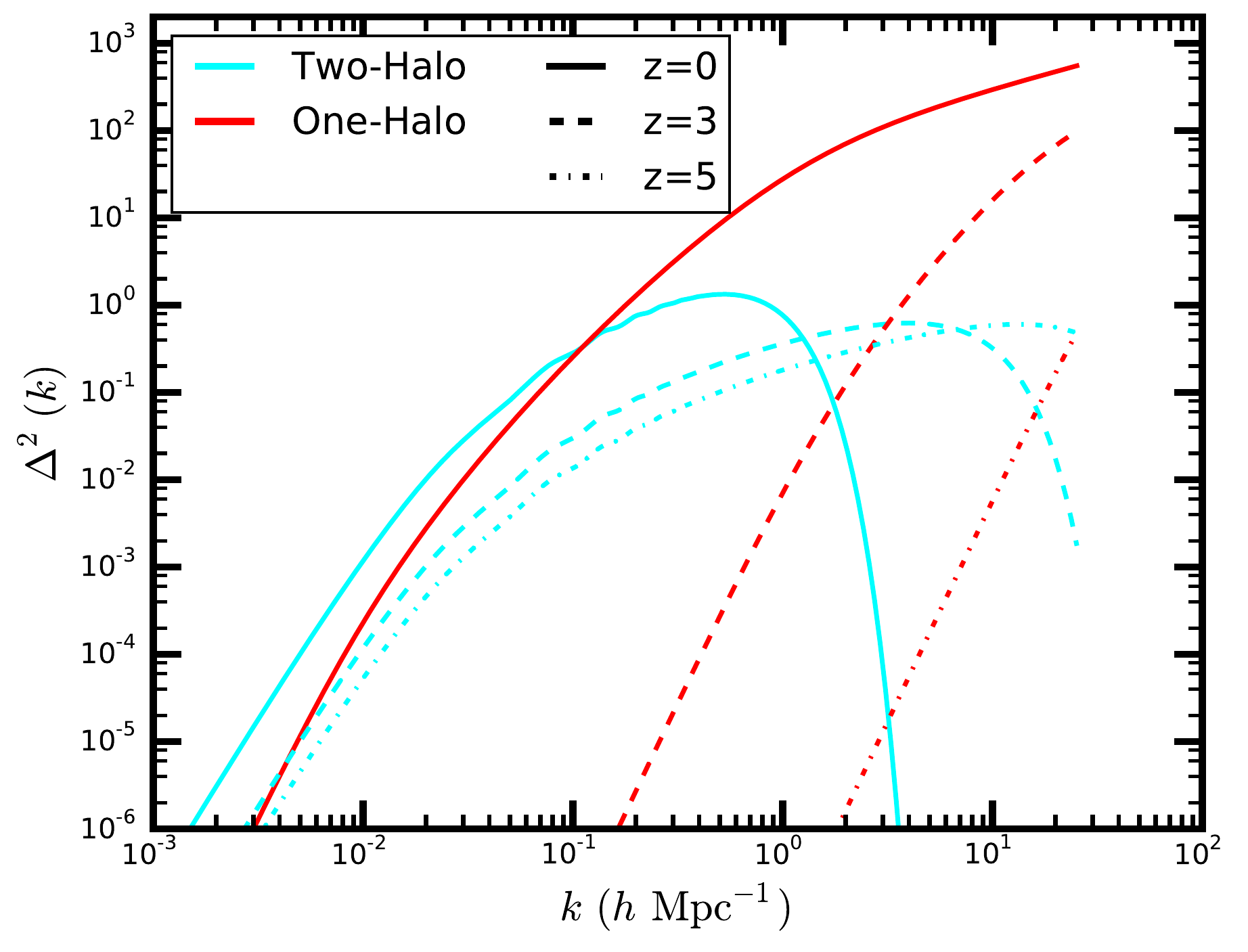}
	\caption{\textit{Left panel}: Time evolution of the linear (orange lines) and non linear (blue lines) power spectra, within the two-halo model considered in this work. The power spectra are drawn at redshifts $z=0$ (solid lines), $z=3$ (dashed lines) and $z=5$ (dot-dashed lines). \textit{Right panel}: Time evolution of the two-halo (cyan lines) and one-halo (red lines) terms, within the two-halo model considered in this work. The functions plotted refer to redshifts $z=0$ (solid lines), $z=3$ (dashed lines) and $z=5$ (dot-dashed lines).}\label{fig:time_evolution}
\end{figure*}

The time evolution of the power spectrum is shown in figure \ref{fig:time_evolution}. In the left panel, I show the linear (orange lines) and non linear (blue lines) power spectra at redshifts $z=5$, $z=3$ and $z=0$, represented with dot-dashed, dashed and solid lines respectively. The linear power spectrum has always the same shape, only the amplitude being affected by time evolution. This is a direct consequence of linear theory, according to which density perturbations are simply rescaled by the growth factor, $\delta(z)=(D(z)/D(z_0))\delta(z_0)$. At $z=5$, the non linear power spectrum is barely distinguishable from the linear one. This means that the scales in the dynamic range shown ($k \lesssim 20 \, h \, \rm Mpc^{-1}$) are still growing linearly. As the redshift decreases, larger and larger scales enter the horizon, so one can see a departure of the non-linear power spectrum from the one predicted by linear theory. The point discriminating between the two regimes moves towards larger scales. The right panel of figure \ref{fig:time_evolution} shows the time evolution of the two-halo (cyan lines) and one-halo (red lines) terms. The line styles have the same meaning as in the left panel. The two-halo term evolves linearly at small $k$ and becomes smaller and smaller at large $k$ as time evolves. On the contrary, the time growth of the one-halo term is predominant at large $k$. Although the one-halo term does present an evolution at small $k$ as well, it is always negligible with respect to the two-halo term.

In developing the generalisation of FKP, I shall assume that the power spectrum can be written as in equation \eqref{eq:two-halo-P}. I shall then disentangle time-evolution effects, providing an estimator for the liner power spectrum at fixed time.

\subsection{Computing Weighted Density Fluctuations} \label{computation F k}

Following the FKP treatment, I define
\begin{equation}
	F(\textbf{k}) =\\
	 \frac{1}{A(\textbf{k})} \int d^3r \, e^{i \textbf{k} \cdot \textbf{r}} w(\textbf{r}, \, \textbf{k}) [n_g(\textbf{r}) - \alpha n_s(\textbf{r})]  
\end{equation}
where $A(\textbf{k})$ is the normalisation function
\begin{equation}
	A(\textbf{k})= \left[ \int  d^3r \, \bar{n}(\textbf{r})^2 w(\textbf{\textbf{r}}, \, \textbf{\textbf{k}})^2 \right]^{\frac{1}{2}}
\end{equation}
I have explicitly included the possible $\textbf{k}$-dependence of the weight function. Taking the expectation value of $\vert F(\textbf{k})\vert^2$ and exploiting the same relations used in the FKP treatment, I obtain
\begin{align*}
	\langle \vert F(\textbf{k})\vert^2 \rangle &= \frac{1}{A(\textbf{k})^2} \int d^3r \int d^3r' e^{i\textbf{k} \cdot (\textbf{r}-\textbf{r}')} w(\textbf{r},\, \textbf{k}) w(\textbf{r}', \, \textbf{k})\bar{n}(\textbf{r}) \bar{n}(\textbf{r}') \xi(\textbf{r}, \, \textbf{r}') \\ &+ \frac{1+ \alpha}{A(\textbf{\textbf{k}})^2} \int d^3r \, w(\textbf{r}, \textbf{k})^2 \bar{n}(\textbf{r})
\end{align*} 
\begin{equation}
	\left.\right.
\end{equation}
The second term is the shot noise, which I shall write as $P_{\textrm{shot}}(\textbf{k})$ from now on. The first term contains the two-point correlation function. One can assume that in a realistic situation the correlation function approaches zero very rapidly as the distance between $\textbf{r}$ and $\textbf{r}'$ increases. Within this assumption and the distant observer approximation, one can write ~\cite{Yamamoto_2003}
\begin{equation}
	\xi(\textbf{r}-\textbf{r}', \, r^*) \approx \int \frac{d^3k}{(2 \pi)^3} \, e^{-i \textbf{k} \cdot (\textbf{r}-\textbf{r}')}\, P(\textbf{k}, \, r^*)
\end{equation}
where $\textbf{r}^*=(\textbf{r}+\textbf{r}')/2$. In other words, the past light-cone effect is encoded by the average distance between the two points considered. One can now plug this relation in the expression for $\langle \vert F(\textbf{k}) \vert^2 \rangle$. It yields
\begin{align*}
	\langle \vert F(\textbf{k}) \vert^2 \rangle &= \frac{1}{A(\textbf{k})^2} \int d^3r \int d^3r' e^{i\textbf{k} \cdot (\textbf{r}-\textbf{r}')} w(\textbf{r},\, \textbf{k}) w(\textbf{r}', \, \textbf{k})\bar{n}(\textbf{r}) \bar{n}(\textbf{r}') 
	\int \frac{d^3k'}{(2 \pi)^3} \, e^{-i \textbf{k}' \cdot (\textbf{r}-\textbf{r}')}\, P(\textbf{k}', \, r^*)\\ &+ P_{\textrm{shot}}(\textbf{k}) .
\end{align*}
\begin{equation}
\label{eq:NL_method}
\left. \right.
\end{equation}
Recalling that 
\begin{equation}
\label{eq:split_power}
	P(\textbf{k}', \, r^*) = \frac{2 \pi ^2}{k'^3} \Delta^2(\textbf{k}', \, r^*) = \frac{2 \pi ^2}{k'^3} [\Delta^2_{\rm Q}(\textbf{k}', \, r^*)+\Delta^2_{\rm H} (\textbf{k}', \, r^*)]\, ,
\end{equation}
the right hand side of equation \eqref{eq:NL_method} can be split so that one can write 
\begin{equation} \label{eq:F_Q_H}
\langle \vert F(\textbf{k}) \vert^2 \rangle - P_{\textrm{shot}}(\textbf{k})= T(\textbf{k})+H(\textbf{k}) \, ,
\end{equation}
where $T(\textbf{k})$ and $H(\textbf{k})$ contain only the two-halo and one-halo terms, respectively. In the next two sections I shall treat each term separately. The goal is to find a strategy to obtain an estimator for the linear power spectrum at fixed time $\Delta^2_{\rm L 0} (\textbf{k})$.

\subsection{Two-halo Term} \label{sec:two-halo term}

The two-term proposed by \cite{Takahashi_2012} is given by equation \eqref{eq:Takahashi_DQ}. Rather than developing an \textit{ad hoc} method for the revised halofit model proposed by \cite{Takahashi_2012}, I shall consider a wider class of models. Specifically, I assume the relation between $\Delta_{\textrm{Q}}^2(k)$ and $\Delta_{\textrm{L}}^2(k)$ to be
\begin{equation}\label{3}
	\Delta^2_{\rm Q} (\textbf{k}, \,r)=h[ \Delta^2_{\textrm{L}}(\textbf{k}),\, \textbf{k}, \, r]= f \left[ \left( \frac{D(r)}{D(r_0)} \right)^2 \Delta^2_{\textrm{L\,0}}(\textbf{k}), \, \textbf{k} \right] g(\textbf{k},\, r)^2
\end{equation}
I have not assumed that $\Delta^2_{\rm Q}$ is a function of the norm of $\textbf{k}$ only, to keep the treatment as general as possible.
I had to include a dependence of $\Delta^2_{\rm Q}$ on $r$ to make sense of the time evolution of the perturbations. In other words, equation \eqref{3} says that the mapping can be written as a product of a function of $\Delta^2_{\textrm{L}}(\textbf{k})$ and $\textbf{k}$ only, and a positive function of $\textbf{k}$ and $\textbf{r}$ only. The mapping expressed in equation \eqref{eq:Takahashi_DQ} falls into this category. 

If the two-halo term has the shape described in equation \eqref{3}, it is possible to re-write $T$ in \eqref{eq:F_Q_H} simply as a polynomial function of the linear power spectrum at fixed time. To make notation lighter, let us set $x=\Delta^2_{\textrm{L}\,0}(\textbf{k})$, so that the function $f$ in \eqref{3} can now be written as $f \left[ ( D(r)/D(r_0) )^2 x, \, \textbf{k} \right]$. Keeping $r$ and $\textbf{k}$ fixed, a different shape of the power spectrum will yield a different value of $x$. For any fixed $\textbf{k}$, one can then Taylor expand the function $f$ in equation \eqref{3} around $x=0$, till a certain order $m$. For $x \ll 1$, the expansion recovers the linear regime. Forcing slightly the aforementioned approximation according to which the correlation function goes to zero rapidly as $\textbf{r}-\textbf{r}'$ increases, one can assume $ r \approx r' \approx r^* $. Equation \eqref{eq:NL_method} then simplifies as
\begin{align*}
	T(\textbf{k}) &\approx \frac{1}{A(\textbf{k})^2} \sum_{j=0}^m \frac{1}{j!} \int \frac{d^3k'}{(2 \pi)^3} \, \frac{2 \pi ^2}{k'^3} \left. \frac{\partial ^j }{\partial x^j}f(x,\,\textbf{k}') \right \vert _{(0, \, \textbf{k}')} x^j \int d^3r \, e^{i \textbf{r} \cdot (\textbf{k} - \textbf{k}')} w(\textbf{r}, \textbf{k}) \bar{n}(\textbf{r}) g(\textbf{k}', r) \left( \frac{D(r)}{D(r_0)} \right)^j\\
	&\int d^3r' \, e^{-i \textbf{r}' \cdot (\textbf{k} - \textbf{k}')} w(\textbf{r}', \textbf{k}) \bar{n}(\textbf{r}') g(\textbf{k}', r') \left( \frac{D(r')}{D(r_0)} \right)^j  \, .
\end{align*}
Defining
\begin{equation}
	G_j(\textbf{k},\, \textbf{k}') = \frac{1}{A(\textbf{k})} \int d^3r \, e^{i \textbf{r} \cdot (\textbf{k} - \textbf{k}')} w(\textbf{r}, \textbf{k}) \bar{n}(\textbf{r}) g(\textbf{k}', r) \left( \frac{D(r)}{D(r_0)} \right)^j \,
\end{equation}
one obtains
\begin{equation} 
	T(\textbf{k}) \approx \sum_{j=0}^m \frac{1}{j!} \int \frac{d^3k'}{(2 \pi)^3} \, \vert G_j(\textbf{k}, \,\textbf{k}') \vert ^2  \frac{2 \pi ^2}{k'^3} \left.\frac{\partial ^j }{\partial x^j}f(x,\,\textbf{k}') \right \vert _{(0, \, \textbf{k}')} x^j \, .
\end{equation}
If the functions $G_j(\textbf{k},\, \textbf{k}')$ are narrow in $k$-space, that is, peaked around $\textbf{k}$ and such that their width is $\sim 1/L$, where $L$ is the depth of the survey, then one can define
\begin{equation}
	I_j(\textbf{k}) = \int \frac{d^3k'}{(2 \pi)^3} \, \vert G_j(\textbf{k}, \textbf{k}') \vert ^2
\end{equation}
and write
\begin{equation}
\label{eq:Q_result}
	T(\textbf{k}) \approx  \sum_{j=0}^m \frac{1}{j!} I_j(\textbf{k}) \frac{2 \pi ^2}{k^3} \left.\frac{\partial ^j }{\partial x^j}f(x,\,\textbf{k}) \right \vert _{(0, \, \textbf{k})} x^j \, .
\end{equation}

\subsection{One-Halo Term}

The one-halo term, as defined by \cite{Takahashi_2012}, cannot be factorised as in equation \eqref{3}. Moreover, it is not a function of $\Delta^2_{\rm L \, 0}$ only, which makes it hard to disentangle the dependence on $r$ and expand it in a Taylor series as it has been done for the two-halo term. The one-halo term depends on $\Delta^2_{\rm L \, 0}$ only through $k_{\sigma}$. Since the linear power spectrum depends on the time $t(r)$ considered, the $r$-dependence of the one-halo term is also encapsulated in $k_{\sigma}$. It is then necessary to introduce an approximation to drop the dependence on $r$. For this purpose, I propose the following ansatz. I replace $\Delta^2_{\rm H}(\textbf{k},\, r)$ in equation \eqref{eq:split_power} with its average over $r$, weighted with the number density of galaxies, that is
\begin{equation}
	\widetilde{\Delta}^2_{\rm H} (\textbf{k}) = \frac{\int d^3r \, \bar{n}(\textbf{r}) \, \Delta^2_{\rm H}(\textbf{k},\,r)}{\int d^3r \, \bar{n}(\textbf{r})} \, .
\end{equation}
In other words, $\widetilde{\Delta}^2_{\rm H} (\textbf{k})$ can be seen an ``effective one-halo term''. The one-halo term $\Delta^2_{\rm H}(\textbf{k},\, r)$ contributes to $\widetilde{\Delta}^2_{\rm H} (\textbf{k})$ at any value of $r$. The contribution is bigger where the number density of galaxies is larger.

Although applied to a different context, this approach is similar to considering a space-averaged estimator of the power spectrum, as in \cite{Yamamoto_2003} (discussed at the end of subsection \ref{sec:final_weighting}). However, in this work only the one-halo term is averaged, and not the entire power spectrum. 

Adopting the same approximations that in the previous paragraph led to equation \eqref{eq:Q_result}, one obtains
\begin{equation}
	H(\textbf{k}) \approx \int \frac{d^3k'}{(2 \pi)^3} \,  \frac{2 \pi ^2}{k'^3} \widetilde{\Delta}^2_{\rm H} (\textbf{k}') \vert L (\textbf{k},\,\textbf{k}') \vert ^2 \, ,
\end{equation}
where I defined
\begin{equation}
	L (\textbf{k},\,\textbf{k}') = \frac{1}{A(\textbf{k})} \int d^3r \, e^{i \textbf{r} \cdot (\textbf{k}-\textbf{k}')} w(\textbf{r},\,\textbf{k}) \bar{n}(\textbf{r}) \, .
\end{equation}
Following what has been done for the $G_j(\textbf{k},\,\textbf{k}')$ functions when computing $T(\textbf{k})$, one can assume that also $L (\textbf{k},\,\textbf{k}')$ is narrow in $k$-space. Under such approximation:
\begin{equation}
\label{eq:H_result}
	H(\textbf{k}) \approx \frac{2 \pi ^2}{k^3}  \widetilde{\Delta}^2_{\rm H} (\textbf{k}) 
\end{equation}
\subsection{Estimator and Optimal Weighting}
\label{sec:final_weighting}

Using equations \eqref{eq:Q_result} and \eqref{eq:H_result}, equation \eqref{eq:NL_method} can be re-written as
\begin{equation}
 \label{nonlinear FKP equation}
	\langle \vert F(\textbf{k}) \vert^2 \rangle - P_{\textrm{shot}}(\textbf{k}) = T(\textbf{k})+H(\textbf{k}) \approx \frac{2 \pi ^2}{k^3} \left( \sum_{j=0}^m \frac{1}{j!} I_j(\textbf{k})  \left.\frac{\partial ^j }{\partial x^j}f(x,\,\textbf{k}) \right \vert _{(0, \, \textbf{k})} x^j +
	   \widetilde{\Delta}^2_{\rm H} (\textbf{k}) \right)\, .
\end{equation}

I define $x$ to be the estimator of the dimensionless linear power spectrum at present time. It can be determined by solving the above equation, for every value of $\textbf{k}$ of interest. The estimator can then be averaged over a shell of volume $V_k$ in Fourier space. I shall indicate the averaged estimator of the linear power spectrum at present time as $\widehat{P}_{\rm L \, 0} (k)$, and $\widehat{\Delta^2}_{\rm L \, 0}(k)$ its dimensionless counterpart. Equation \eqref{nonlinear FKP equation} is the main result of this work.

The variance of the estimator is given by	$\sigma ^2 _{P_{\rm L \, 0}} (\textbf{k}) = \langle (\widehat{P}(\textbf{k})-P(\textbf{k}))^2 \rangle$. To compute the variance, I shall follow the same reasoning and approximations adopted in FKP, using the identity
\begin{equation}
\label{eq:identity}
\langle (\widehat{P}_{\rm L \, 0}(\textbf{k})-P_{\rm L \, 0}(\textbf{k})) (\widehat{P}_{\rm L \, 0}(\textbf{k}')-P_{\rm L \, 0}(\textbf{k}')) \rangle = \vert \langle F(\textbf{k}) F^*(\textbf{k}') \rangle \vert ^2 \, .
 \end{equation}
Considering $\textbf{k}'=\textbf{k}+\delta \textbf{k}$, with $\vert \delta \textbf{k} \vert \ll k$, one obtains
\begin{equation}
 \langle F(\textbf{k}) F^*(\textbf{k}+\delta \textbf{k}) \rangle \approx \frac{2\pi^2}{k^3} \sum _j \frac{1}{j!} \left. \frac{\partial ^j }{\partial x ^j} f(x,\,\textbf{k}) \right \vert _{(0, \textbf{k})} x^j Q_j(\delta \textbf{k}, \,\textbf{k}) + R(\delta \textbf{k},\, \textbf{k}) \frac{2\pi^2}{k^3} \widetilde{\Delta}^2_{\rm H}(\textbf{k}) + S(\delta \textbf{k},\, \textbf{k}) \, ,
\end{equation}
where
\begin{align}
Q_j(\delta \textbf{k},\, \textbf{k})  &= \frac{1}{A(\textbf{k})^2} \int d^3r \, e^{i \textbf{r} \cdot \delta \textbf{k}} \, \left(\frac{D(r)}{D(r_0)} \right)^{2j} g(\textbf{k}, r)^2  \bar{n}(\textbf{r})^2 w(\textbf{r},\,\textbf{k})^2\\
R(\delta \textbf{k}, \,\textbf{k})  &= \frac{1}{A(\textbf{k})^2} \int d^3r \, e^{i \textbf{r} \cdot \delta \textbf{k}}  \, \bar{n}(\textbf{r})^2 w(\textbf{r},\,\textbf{k})^2\\
S(\delta \textbf{k}, \,\textbf{k})  &= \frac{1+\alpha}{A(\textbf{k})^2} \int d^3r \, e^{i \textbf{r} \cdot \delta \textbf{k}} \,  \bar{n}(\textbf{r}) w(\textbf{r},\,\textbf{k})^2 \, .
\end{align}
The variance of the estimator, averaged in a spherical shell of volume $V_k$ in Fourier space, is then given by 
\begin{equation}
\sigma ^2 _{P_{\rm L \, 0}} (k) \approx \frac{1}{V_k} \int _{V_k} d^3 k' \, \left \vert \frac{2\pi^2}{k^3} \sum _j \frac{1}{j!} \left. \frac{\partial ^j }{\partial x ^j} f(x,\,\textbf{k}) \right \vert _{(0, \textbf{k})} x^j Q_j(\delta \textbf{k}, \,\textbf{k}') + R(\delta \textbf{k},\, \textbf{k}') \frac{2\pi^2}{k^3} \widetilde{\Delta}^2_{\rm H}(\textbf{k})  + S(\delta \textbf{k},\, \textbf{k}')  \right \vert ^2 \, .
\end{equation}
In the limit $\alpha \rightarrow 0$, it can be shown that
\begin{align*}
\frac{\sigma ^2 _{P_{\rm L \, 0}} (k)}{P_{\rm L \, 0} (k)^2} &\approx \frac{(2\pi)^3}{V_k A(k)^4} \int d^3r \, \bar{n}(\textbf{r})^4 w(\textbf{r},\,k)^4 \left[ \frac{1}{\bar{n}(\textbf{r}) P_{\rm L \, 0} (k)}\right.\\ &+\left. \frac{1}{P_{\rm L \, 0} (k)} g(k,\,r)^2 \frac{2\pi^2}{k^3} \sum _j \frac{1}{j!}  \left. \frac{\partial ^j }{\partial x ^j} f(x,\,k) \right \vert _{(0, k)} x^j  \left(\frac{D(r)}{D(r_0)} \right)^{2j} \right. \\ &+ \left. \frac{1}{P_{\rm L \, 0} (k)} \frac{2\pi^2}{k^3}  \widetilde{\Delta}^2_{\rm H}(k) \right]^2\\
\end{align*}
\begin{equation}
\left. \right.
\end{equation}
To obtain the optimal weight function $w_0(\textbf{r})$, I require the above expression to be stationary upon arbitrary variations $\delta w(\textbf{r})$ of the weight function. The result is
\begin{equation} 
\label{non linear weight function}
	w_0(\textbf{r}, k) = \frac{P_{\rm L \,0}(k)}{1 + \bar{n}(\textbf{r}) \frac{2 \pi ^2}{k^3} \left[ g(k,\,r)^2  \sum _j \frac{1}{j!}  \left. \frac{\partial ^j}{\partial x^j} f(x,\,k) \right\vert _{(0, \, k)} x^j \left( \frac{D(r)}{D(r_0)} \right)^{2j}+\widetilde{\Delta}^2_{\rm H}(k) \right]} = \frac{P_{\rm L\, 0}(k)}{1 + \bar{n}(\textbf{r}) P(k,\, r)} \, .
\end{equation} 
Expanding the denominator the weight function in the above expression only to first order, equation \eqref{eq:weight_func_linear} is recovered. As a caveat, I recall that the identity \eqref{eq:identity} holds only if $F(\textbf{k})$ is Gaussian distributed. In the strongly non-linear regime the distribution of density fluctuations is not Gaussian, so in principle the above weight function may not be valid for any $k$. Nevertheless, the assumption of Gaussianity is valid at large scales ~\cite{FKP}, so the ``real'' optimal weight function should coincide with equation (\ref{non linear weight function}) at these scales. Therefore, even if (\ref{non linear weight function}) is not the optimal weight function, one expects it to be close to the actual one, by continuity. Furthermore, I will show in section \ref{sec:implementation} that the method is accurate up to the mildly non-linear regime ($\Delta^2 (k) \gtrsim 1$), so in practice there is no reason to worry about the strongly non-linear regime ($\Delta^2 (k) \gg 1$).

The expression of the weight function at the right hand side of the last equality in equation \eqref{non linear weight function} is consistent with the findings by \cite{Yamamoto_2003}. However, the results in \cite{Yamamoto_2003} are only formally identical to the ones presented here. Indeed, the estimator defined by \cite{Yamamoto_2003} is
\begin{equation} \label{Yamamoto_estimator}
	\widetilde{P}(\textbf{k}) = \frac{1}{A^2} \int d^3r \, \bar{n}(\textbf{r})^2 w(\textbf{r})^2  P(\textbf{k},\, r) \, .
\end{equation}
This quantity does not estimate the power spectrum at fixed time, but the power spectrum integrated over a certain range in $r$. Thus, it is different from $\widehat{P}_0(k)$. In \cite{Yamamoto_2003} no particular functional shape describing the growth of perturbations is assumed. In addition to the light-cone, also other effects, like redshift space distortions and bias, are included in \cite{Yamamoto_2003}. Consequently, the result in \cite{Yamamoto_2003} is more general, but provides an estimator of the power spectrum integrated over a certain redshift range. On the other hand, the method presented in this manuscript is more model-dependent, but provides an estimator of the power spectrum at fixed time, disentangling the growth of perturbations. Both methods are correct, as long as one clearly bears in mind which estimator was defined, and is coherent with that definition. 

\section{Implementation of the Method}
\label{sec:implementation}

I shall now outline the steps of the code that has been implemented to test the method, underscoring the approximations adopted.

First of all, I choose the cosmological parameters given by \cite{Planck_2014} ($\Omega_{\mathrm{m}}=0.315$, $\Omega_{\Lambda}=1-\Omega_{\mathrm{m}}=0.683$, $\Omega_{\mathrm{b}}=0.049$, $h=0.673$, $A_s=2.215 \times 10^{-9}$, $n_s=0.96$) as the fiducial cosmological model. I then generate the corresponding linear power spectrum at redshift $z=0$ using software CAMB ~\cite{CAMB_1, CAMB_2}. I assume it to be the true power spectrum. Density fluctuations are assumed to be an isotropic random field, so the power spectrum depends only on the norm of $\textbf{k}$.

\begin{figure}
	\includegraphics[width=0.49\columnwidth]{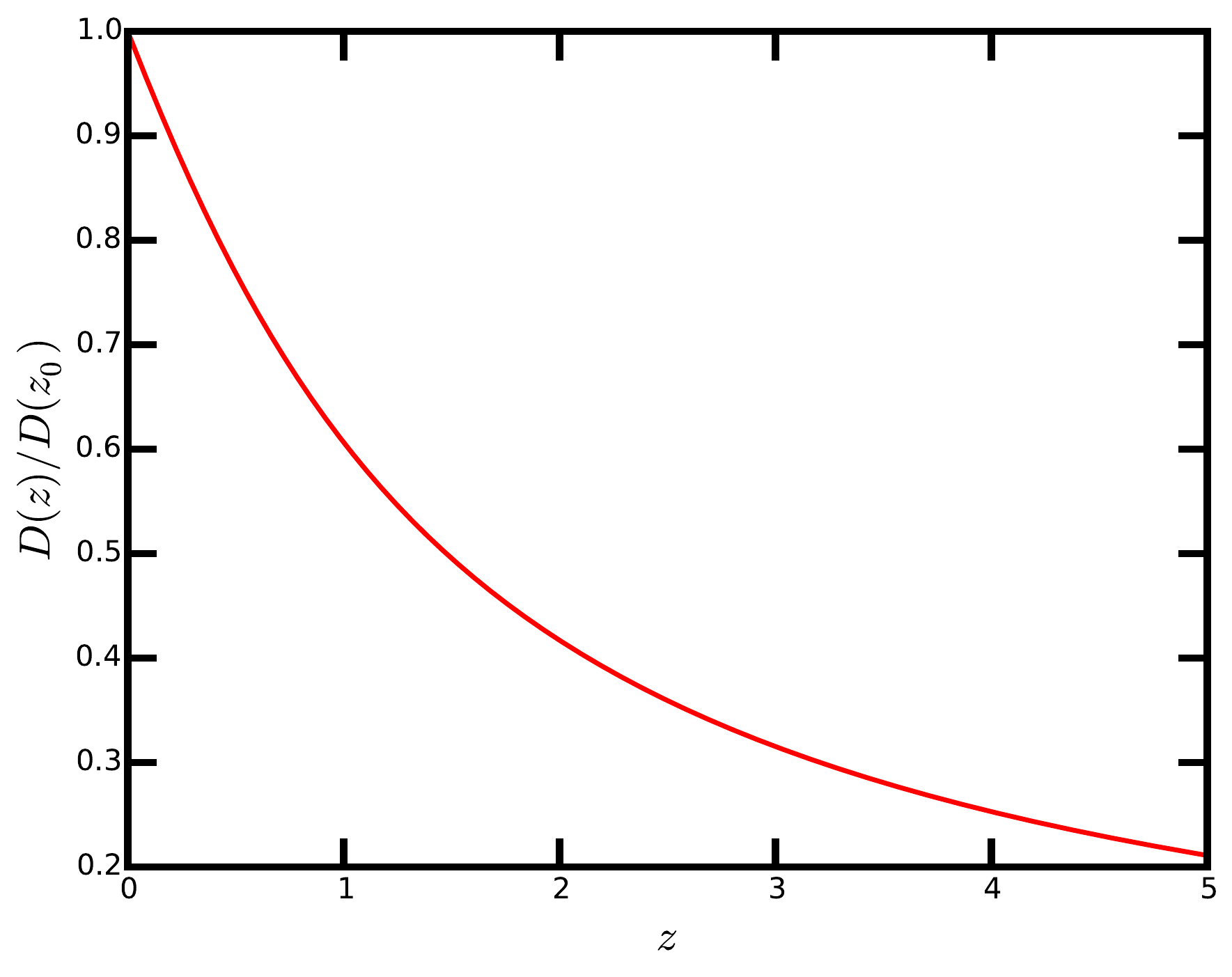}
	\includegraphics[width=0.49\columnwidth]{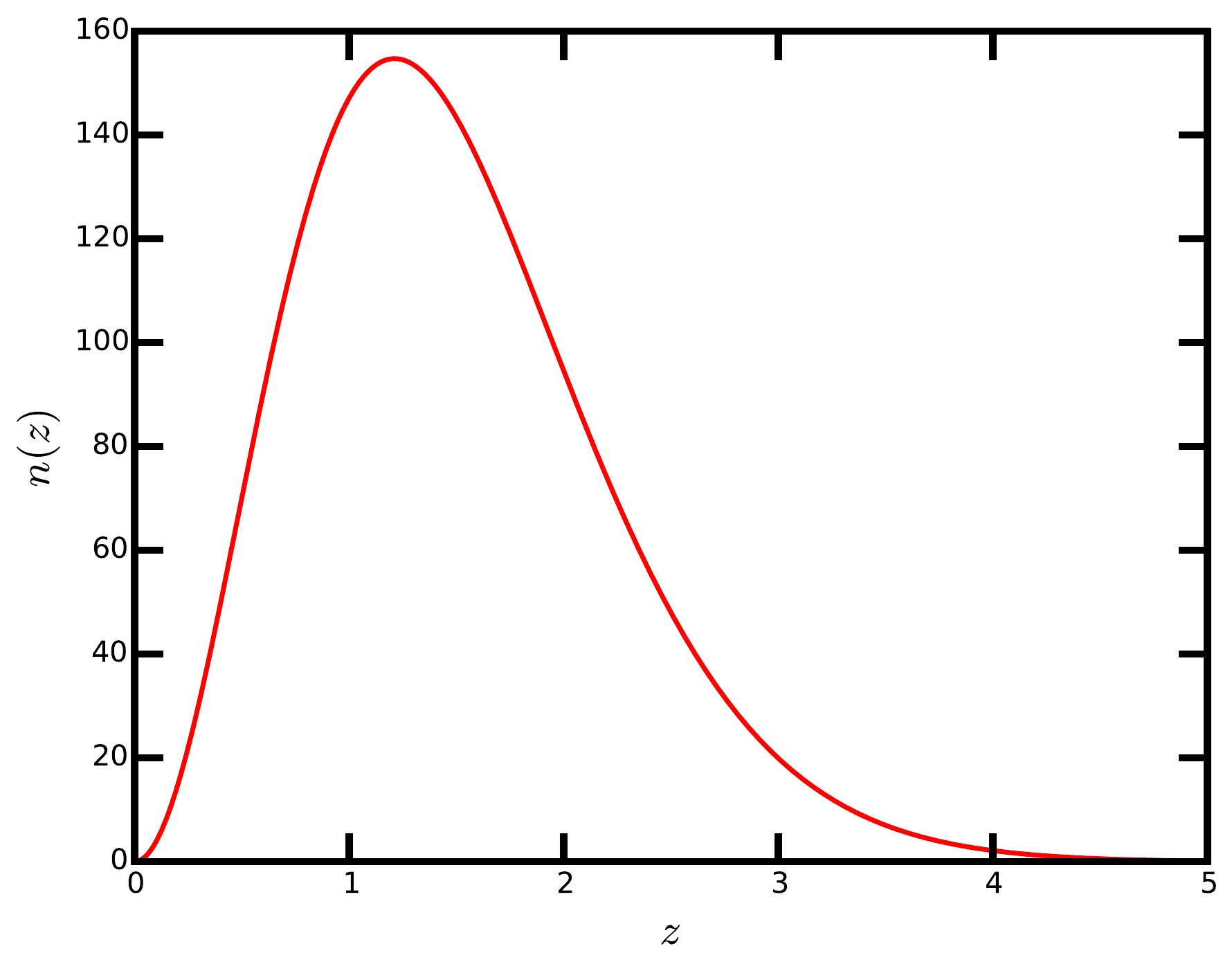}
	\caption{\textit{Left panel}: Growth function versus redshift, for the cosmological model and parameters governing the growth of perturbations considered. The growth functions are normalised to the redshift $z_0$, at which the linear power spectrum is estimated. \textit{Right panel}: Redshift-dependence of the number density of galaxies assumed in the test of the method presented in this work.} \label{fig:growth_nz}
\end{figure}

Then, the code computes the redshift-distance relation for the chosen cosmology and thus is able to obtain the growth function $D(r)$ using 
\begin{equation}
	\label{eq:growth_eq}
	\frac{d \ln \delta (z)}{d \ln a(z)}=\Omega _m ^{\gamma} (z) \, ,
\end{equation} 
where $\gamma$ is a constant depending on the theory of gravitation considered. For General Relativity, its value is 0.55 ~\cite{Euclid}. The growth function is plotted in the left panel of figure \ref{fig:growth_nz}.

Functions $\sigma ^2 (R)$ and $k_{\sigma}(r)$ are then computed, as well as all the quantities needed to define the one- and two-halo terms according to \cite{Takahashi_2012}. In this particular model, the function $f(\Delta^2_{\textrm{L}}(k),\,k)$ defined in section \ref{sec:two-halo term} actually depends only on the norm of $k$ and only through $\Delta^2_{\textrm{L}}(k)$. When one considers time evolution, there is of course an additional $r$ dependence. Regarding the function $g(k,\,r)$ defined in section \ref{sec:two-halo term}, the $r$-dependence is encoded by the function $k_{\sigma}(r)$. 

Now one has to define the number density and the weight function. Regarding the number density, I assume a realistic shape given by \cite{Smail_1994}
\begin{equation}
	\bar{n}(z) = C z^2 \exp \left[ - \left( \frac{z}{\widetilde{z}} \right) ^{\frac{3}{2}} \right] 
\end{equation}
I chose $\widetilde{z}=1$ and $C=400$. The corresponding shape of $n(z)$ can be seen in the right panel of figure \ref{fig:growth_nz}.
The dependence of the number density on $r$ is then easily found using the redshift-distance relation previously computed. I then adopted the weight function minimising the variance.

One can compute $\langle \vert F(\textbf{k}) \vert ^2  \rangle -P_{\textrm{shot}}(\textbf{k})$ using \eqref{nonlinear FKP equation}. Since all quantities depend now only on the norm of $r$ and $k$, the expressions for $G_j(k, k')$ and $I_j(k)$ are simplified accordingly. 

Using all the expressions computed so far, equation \eqref{nonlinear FKP equation} is solved for different values of $k$ and at different orders of the expansion, restricting the solutions to real numbers. In this way, one can see if the equation admits real solutions and, if so, how many. In case more than one real solution is possible, one can try to identify the correct one through physical arguments. For example, I found out that one of the solutions of equation \eqref{nonlinear FKP equation} was negative for all $k$. The power spectrum is by definition positive, so one could clearly identify the physical solution. One can use such criterion at any order. Even if there are two positive solutions for all $k$ at a certain order $m$, one can still understand which is the correct one by considering that, at small $k$, it should not deviate appreciably from the solution at order $m-1$.  I can then investigate the validity of the method with further analysis, comparing the solutions at different orders with the power spectrum provided by CAMB. 

Before showing the results, let us point out one last remark. In all the quantities considered, integrals over $r$ and $k$ are extended over the whole three-dimensional space. Of course, it is not possible to integrate till infinity, so cut-off values must be set. The output of CAMB provides $P(k)$ till a certain wavenumber. The power spectrum tends to zero at large $k$. I then extrapolate $P(k)$ to $k_{\textrm{max}}$, defined so that $P(k_{\textrm{max}})=0$. In other words, instead of considering a power spectrum which tends asymptotically to zero as $k$ tends to infinity, I assume that it is identically zero from a sufficiently large value $k_{\textrm{max}}$ till infinity. Similarly, also $\sigma^2 (R)$ tends to zero at large $R$. I extrapolate again and define $r_{\textrm{max}}$ such that $\sigma^2 (r_{\textrm{max}})=0$. With the cut-off values considered, I am able to span a range in which the growth of perturbations is significant. For example, at $z=3$, where the chosen number density of galaxies is not yet small, the growth factor is about 3.

\section{Results}
\label{Results}

\begin{figure}
\includegraphics[width=\columnwidth]{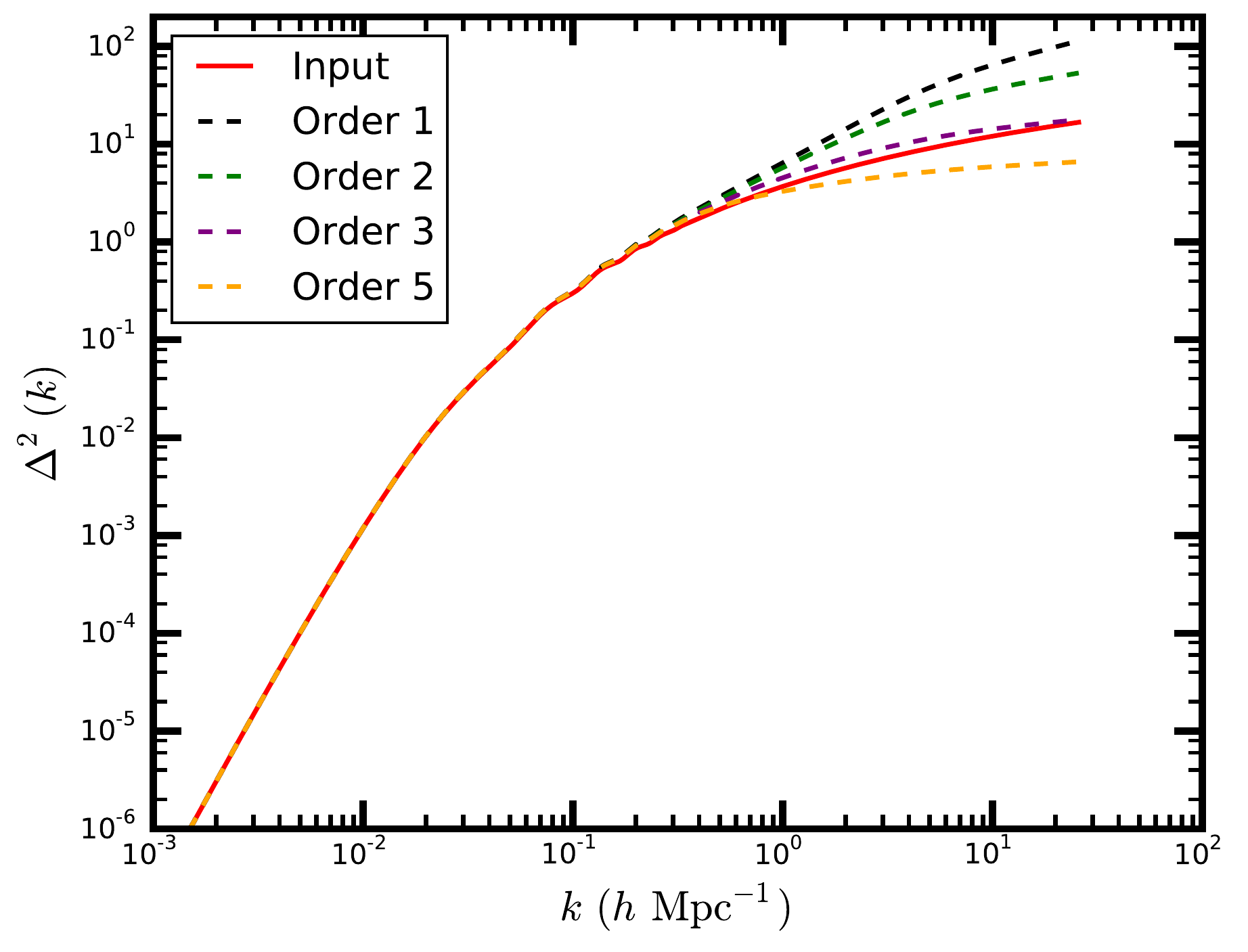}
\caption{Assumed linear power spectrum (solid red line) and its estimators at orders 1, 2, 3 and 5 (solid blue, green, orange, cyan lines, respectively). All quantities are plotted at redshift $z=0$.} \label{Delta L by all}
\end{figure}

In figure \ref{Delta L by all} I show the input linear power spectrum at present (solid red line), together with the solutions of equation \eqref{nonlinear FKP equation} at orders 1, 2, 3 and 5 (dashed black, green, purple and orange lines, respectively).\footnote{It is not so significant to plot the solution at order 4, since it does not admit any real solution for $k \gtrsim 0.44 \, h^{-1} \, \rm Mpc$.} The solutions at all orders considered reproduce very well the input linear power spectrum in the linear regime, as expected. At first order, one can recover $\Delta^2_{\rm L \, 0}$  up to $k \sim 0.1 \, h \, \rm Mpc^{-1}$. Increasing the order, the solutions approach the input linear power spectrum at higher $k$, being able to recover it even in the non linear regime. At order 5 the method seems to approximate $\Delta^2_{\rm L \, 0}$ even up to $k \sim 1 \, h \, \rm Mpc^{-1}$.

\begin{figure}
	\includegraphics[width=\columnwidth]{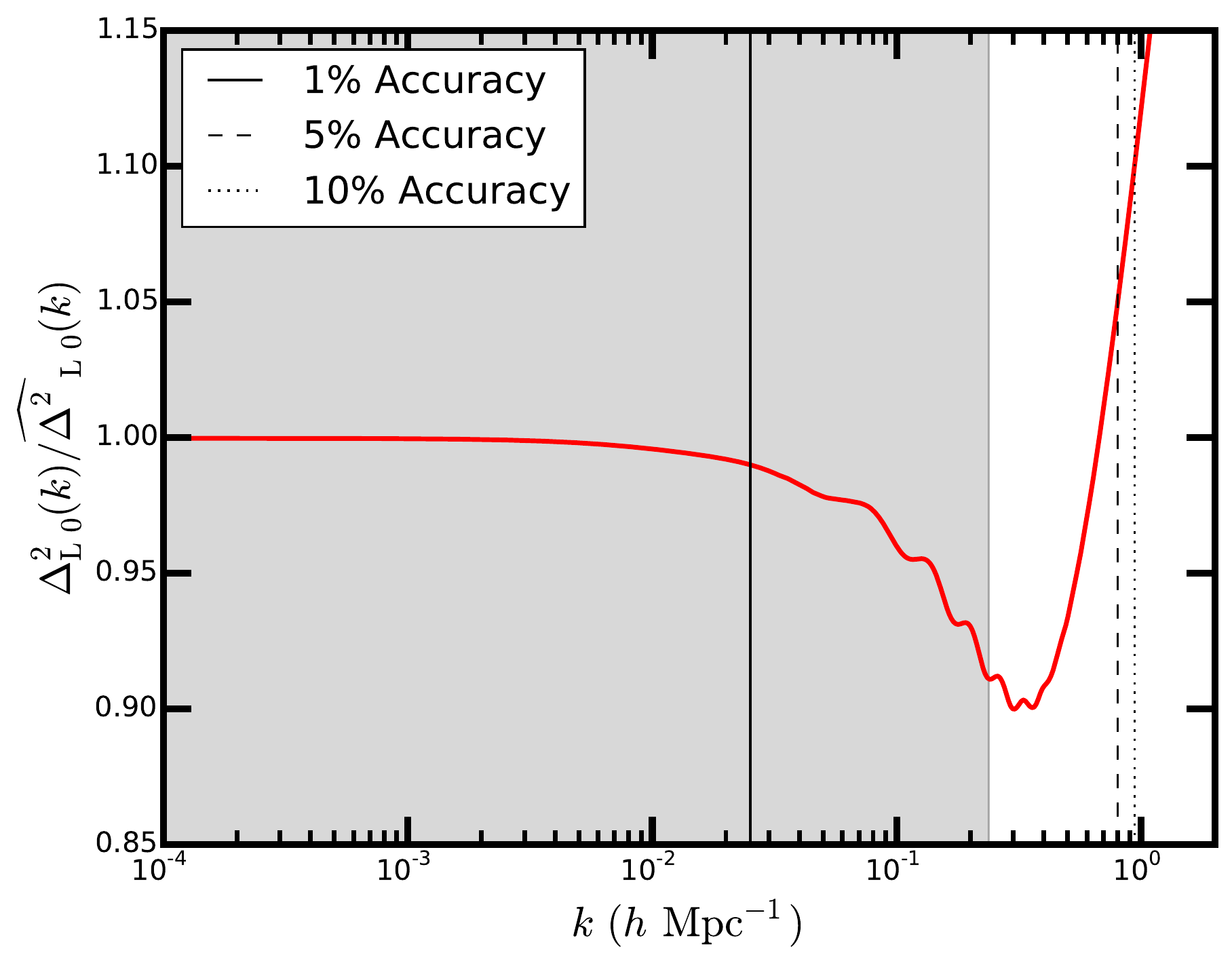}
	\caption{Ratio between the estimator of the linear power spectrum at fifth order and the assumed true linear power spectrum. The quantities considered refer to redshift $z=0$. The vertical lines mark the upper limit of the range of $k$ in which the estimator recovers the true power spectrum within 1\% (solid black line), 5\% (dashed black line) and 10\% (dotted black line) accuracy. The grey shaded area represents the region where the true linear power spectrum at redhsift $z=0$ is smaller than 1, i.e. where linear growth is a legitimate approximation.}\label{Delta L ratio}
\end{figure}

The goodness of the solution obtained at each order is quantified by computing the ratio between $\Delta ^2 _{\rm L \,0}$ and that solution. I then looked for the value of $k$ up to which the ratio deviates from unity within $\pm 0.01$, $\pm 0.05$ and $\pm 0.1$. In other words, I sought the range of $k$  in which the prescribed solution reproduces  $\Delta^2_{\rm L \, 0}$ within $1\%$, $5\%$ and $10\%$ accuracy, respectively. In the following, I shall denote the upper bound of such range as $k_{\rm lim}$. The best result is obtained at fifth order. The ratio between the input power and this solution is plotted in figure \ref{Delta L ratio}. At small $k$, this is very close to 1. As one moves towards smaller scales, the ratio fluctuates, eventually becoming larger and larger. The values of $k_{\rm lim}$ corresponding to the aforementioned levels of accuracy are $0.025 \, h \, \rm Mpc^{-1}$, $0.80\, h \rm Mpc^{-1}$ and $0.94\,h\rm\, Mpc^{-1}$ and are marked with vertical solid, dashed and dotted lines, respectively. The dashed grey area corresponds to the range of $k$ where $\Delta^2_{\textrm{L}\,0} < 1$, that is where linear theory holds. Remarkably, the method presented in this work is able to recover the input linear power spectrum with $5\%$ in a range of modes where non linearities are already significant. 

\begin{figure}
	\includegraphics[width=\columnwidth]{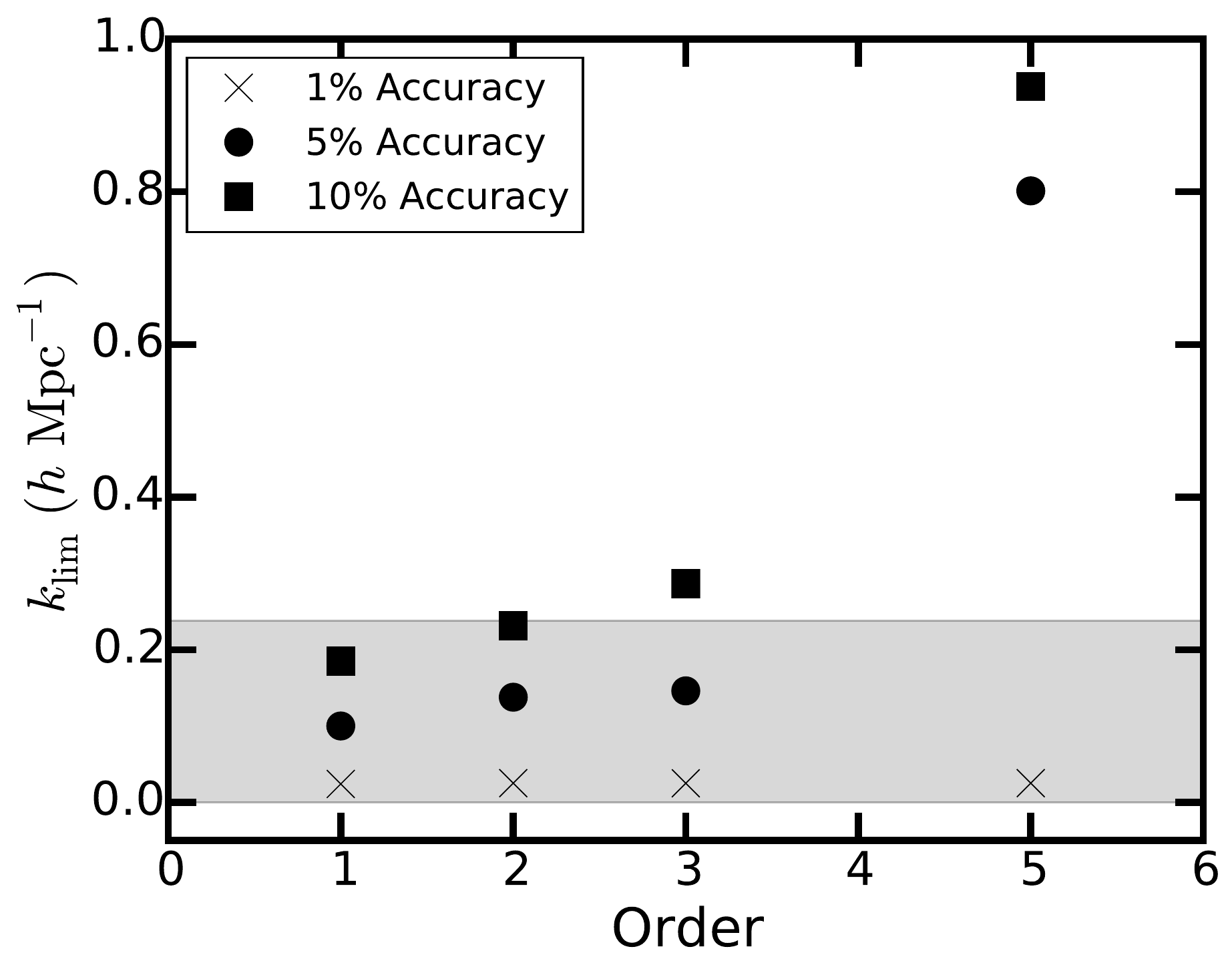}
	\caption{Upper limit of the range in which the estimated linear power spectrum recovers the assumed true linear power spectrum within 1\%, 5\% and 10\% accuracy (black crosses, circles and squares, respectively), as a funciton of the order of the Taylor expansion used to compute the estimator. The results plotted refer to redshift $z=0$. The grey shaded area represents the region where the true linear power spectrum at $z=0$ is smaller than 1, i.e. where linear growth is a legitimate approximation. The values of $k_{\rm lim}$ are positive by definition. The $y$-axis starts from negative values only to make the plot more readable.}\label{fig:klim_order}
\end{figure}

The analysis explained above has been done for all orders considered. The results can be seen in figure \ref{fig:klim_order}. The plot shows $k_{\rm lim}$ as a function of the order of the solution, for $1\%$ (black crosses), $5\%$ (black circles) and $10\%$ (black squares) accuracy. Once again, the shaded grey area indicates the dynamic range where the linear regime is valid. The points at order 4 are missing because there is no real solution beyond $k\sim 0.44\,h \, \rm Mpc^{-1}$. I verified that, beyond fifth order, the range of validity of the method for each level of accuracy does not increase. Therefore,  the solution at order 5 gives the best estimator of $\Delta^2_{\rm L \, 0}$.


\section{Discussion}
\label{Discussion}

The method presented in this work is able to reproduce the fixed-time linear power spectrum of matter density fluctuations in a wide dynamic range. Its range of validity has been determined by considering three levels of accuracy, namely $1\%$, $5\%$ and $10\%$. 

The most restrictive threshold has been chosen because the future Euclid survey is expected to estimate cosmological parameters (which influence the shape of the power spectrum) with a precision of a few percent ~\cite{Euclid}. However, considering possible errors arising from data analysis pipelines based on N-body simulations, 1\% accuracy may be too optimistic. Indeed, simulations showed that baryonic physics may have a non-negligible impact on the matter power spectrum, so ignoring the role of baryons when analysing Euclid-like data could yield systematic errors in the inference of cosmological parameters. 

Different feedback models predict discrepancies up to tens of percent (see e.g. ~\cite{van_Daalen_2011}),  so even a modestly accurate method to infer the matter power spectrum from data could be able to discriminate among different feedback models. Even though at present observational data show no evidence for significant feedback effects on the matter power spectrum at $k \lesssim 2 \, h \, \rm Mpc^{-1}$ \cite{Kitching_2016}, it may still be sensible to expect an accuracy worse than 1\% for Euclid. Furthermore, the accuracy of the halofit model is only around 5\% for $k \sim 1 \, h\, \rm Mpc^{-1}$ at $0<z<3$ ~\cite{Takahashi_2012}, so it makes sense to consider a looser requirement of accuracy for the method presented in this paper. 

For all the aforementioned reasons, a more sensible accuracy threshold to determine the range of validity of the method would be 5-10\%.  I recall that the method presented here reproduces the input power spectrum up to $k\sim 0.80\, h \rm Mpc^{-1}$, already well into the non linear regime, with an accuracy comparable to, or better than, the underlying halofit model. Relaxing the demands on the accuracy down to $10\%$, one can recover the input power spectrum up to almost $1\,h \, \rm Mpc^{-1}$. These are non trivial results, especially if one considers that these performances could make the method applicable to future large-scale surveys like Euclid.

The method allows estimating the linear power spectrum at a fixed time. The estimator is not integrated over a redshift range, so the time evolution of density perturbations has been disentangled. The attempt by \cite{Yamamoto_2003} of dealing with the estimation of the power spectrum in the presence of the light-cone effect is more general, as it does not assume one particular model for the growth of density perturbations. Also, further effects, like redshift space distortions and bias, are considered. On the other hand, it does not completely disentangle time evolution. Thus, in that respect the method presented here is more effective. The two methods are then somewhat complementary to each other. The optimal weight functions obtained in the two cases are consistent.

When implemented, the method presented here is computationally very cheap. The estimation of $\Delta^2_{\textrm{L}\,0}$ is reduced to the calculation of a finite number of integrals and the solution of a polynomial equation. Given sufficient computing time, it is always possible to find the solution. For the levels of accuracy tested, the method achieves its widest range of validity already at order 5. I also tested the procedure with a less realistic trend for the number density of galaxies (i.e. a constant) and still the results are consistent with the ones presented in this manuscript. 

The method has been validated for the halo model, so it is model-dependent.  This issue could be circumvented using forward modeling and statistical inference. One could assume cosmological parameters and a value for $\gamma$ in equation \eqref{eq:growth_eq}. Consequently, one would have the shape of the growth function and the linear power spectrum at present time, obtaining a prediction for $\langle \vert F(k) \vert ^2 \rangle - P_{\textrm{shot}}(k)$. In a real survey, this quantity is obtained from data, so one can get the the likelihood and then, exploiting Bayes theorem, the probability distribution of the parameters, given the data. 

There is still some merit in finding an estimator of $P(k)$ with the strategy outlined in this work, though. An analogy with the CMB may be helpful to go through this point. In the case of the CMB, an enormous amount of data is collected (e.g. \cite{Planck_2014}). The information given by the temperature map is compressed into the angular power spectrum, in a model independent way.\footnote{The only mild assumption is that the distribution of the spherical harmonic coefficients of the temperature distribution is Gaussian.} This statistic actually encodes the information which is necessary to infer the probability of a certain model, given the data. It is not obvious that a similar compression of the data could be done in the case of the power spectrum of matter density fluctuations as well. The method described in this work represents an attempt of doing that. Not surprisingly, it turned out that I had to assume both the growth function and a model describing non linearities. Hopefully, in future work one could be able to relax the assumptions about the latter, making the technique less model-dependent. Regarding the growth function, the situation is different, since it does depend on the cosmological model. One could first of all check how sensitive the method is to a change of the functional form of the growth function. If it turns out that it is dependent on just one parameter, e.g. the derivative of the growth function at some intermediate redshift, one could consider a family of power spectra, each one corresponding to a different value of such parameter, thus providing an estimator for each one of them. So, further insight is needed also in this direction. Another perspective is to generalise the implementation of the method dropping the various approximations described in section \ref{sec:implementation}.

\section{Conclusions and Perspectives}
\label{Conclusions}

I extended the method by \cite{FKP}, incorporating the light-cone effect. I defined a minimum-variance estimator for the linear power spectrum of matter density fluctuations at fixed time and provided and analytic expression for the weight function. This has been done considering non-linear growth of density perturbations, within the context of the halofit model. 

The weight function minimising the variance of the estimator has the same shape as in the original FKP paper, except for the additional dependence of the power spectrum on distance, as shown in equation \eqref{non linear weight function}.  The result presented in this work is formally identical to \cite{Yamamoto_2003}, but has a different physical meaning. Nevertheless, the two methods are consistent. With respect to Yamamoto's method, the technique presented in this work completely disentangles the time evolution of density perturbations, providing an estimator of the linear power spectrum at fixed time. On the other hand, \cite{Yamamoto_2003} takes into account more effects, such as redshift space distortions and bias.

The estimator of the fixed-time linear power spectrum defined in this work is one of the solutions of \eqref{nonlinear FKP equation}, which is the main result of the paper. Equation \eqref{nonlinear FKP equation} relies on a Taylor expansion, meaning that one obtains a different solution depending on the order considered. It is noteworthy that the method reduces to the computation of a finite number of integrals and the solution of a polynomial equation. Given sufficient computing time, a solution can always be found.

The method has been tested for the $\Lambda$CDM cosmology with the parameters given by \cite{Planck_2014}, assuming the revised halofit model by \cite{Takahashi_2012} and a shape for number density of galaxies as in \cite{Smail_1994}. At order 5 of the Taylor expansion in equation \eqref{nonlinear FKP equation}, the estimator of the linear power spectrum at present time reproduces the input power spectrum with $1\%$, $5\%$ and $10\%$ accuracy up to $k\sim 0.025 \, h \, \rm Mpc^{-1}$, $k\sim 0.80 \, h\,\rm Mpc^{-1}$ and $k\sim  0.94 \, h \,\rm Mpc^{-1}$, respectively. The range of validity does not increase at higher orders. 

It is remarkable that the method is capable of deriving the fixed-time linear power spectrum within $5\%-10\%$ well above $k\sim 0.20\,h\,\rm Mpc^{-1}$, where linear theory breaks down ~\cite{Taruya_2012}. This means that the method it potentially useful for analysing the data of future surveys, like Euclid. Indeed, although the expected accuracy of Euclid is 1\% \cite{Euclid}, ignoring baryonic physics in the data analysis may introduce systematic errors, degrading the accuracy. For example, different feedback models could introduce discrepancies in the predicted matter power spectrum up to the order of tens of percent (e.g. \cite{van_Daalen_2011}, but see also \cite{Kitching_2016}).

Despite its simplicity and accuracy over a significant dynamic range, the method presented in this work is model dependent. It assumes that the growth of density perturbations is set by General Relativity and that non linearities in the power spectrum are described by the halo model. Of course, it would be desirable to relax the dependence of the method on any of these assumptions. The former could be addressed by obtaining a family of estimators for different values of the parameter $\gamma$ in \eqref{eq:growth_eq} and compare each one of them with data, to discern which value gives the best fit. Dealing with the latter hypothesis would probably require modeling non linearities with less parameters. 

As an alternative method, one could consider a forward modelling approach. One could parametrise the growth function and the linear power spectrum at present time to predict the left hand side of \eqref{nonlinear FKP equation}. This would be estimated from data of surveys, so that one could apply Bayes theorem to obtain the probability distribution of the parameters, given the data. One could then investigate to which extent the data could be compressed in a model-independent way, although it is not obvious that it can be done.

There are also other ways to improve the method presented in this work. First of all, one could include other effects, like redshift space distortions and bias. Moreover, the technique has been implemented exploiting the fact that the correlation function drops rapidly to zero as the distance between points increases. Nevertheless, \cite{Raccanelli_2014} suggest that this approximation does not hold well at large scales. All the aforementioned assumptions merit further study.





\acknowledgments

Significant parts of this work have been carried out both at the Max Planck Institute for Astronomy in Heidelberg (Germany) and at the  Imperial College of Science, Technology and Medicine in London (United Kingdom). I thank Alan F. Heavens (Imperial College London) and Sabino Matarrese (University of Padua) for their insightful advice and stimulating exchange of ideas. I also thank Giuseppe Tormen (University of Padua) and Joseph F. Hennawi (Max Planck Institute for Astronomy) for helpful comments and discussions. For carrying out this work, I acknowledge financial support by the Galilean School of Higher Studies (Padua, Italy). 


\bibliographystyle{JHEP}
\bibliography{biblio}
%
%
%





\end{document}